\documentclass[11pt]{article}
\usepackage[utf8]{inputenc}
\usepackage{amsmath,amssymb,amsthm,stmaryrd}
\usepackage{xcolor}
\usepackage[margin=1in]{geometry}
\usepackage{hyperref}

\newtheorem{theorem}{Theorem}
\newtheorem{lemma}[theorem]{Lemma}
\newtheorem{proposition}[theorem]{Proposition}

\newtheorem{corollary}[theorem]{Corollary}
\theoremstyle{definition}
\newtheorem{definition}[theorem]{Definition}
\newtheorem{remark}[theorem]{Remark}
\newtheorem{example}[theorem]{Example}

\newcommand{\iso}{\cong}
\newcommand{\Nf}{\mathsf{Nf}}
\newcommand{\Ne}{\mathsf{Ne}}
\newcommand{\mset}[1]{\{\!|\,#1\,|\!\}}
\newcommand{\swap}{\mathsf{swap}}
\newcommand{\HP}{\mathrm{HP}}
\newcommand{\Sub}{\mathsf{Sub}}
\newcommand{\one}{\mathbf{1}}
\newcommand{\two}{\mathbf{2}}
\newcommand{\UA}{\mathrm{UA}}

\newcommand{\sem}[1]{\llbracket #1\rrbracket}
\newcommand{\codeurl}{\url{https://github.com/txa/stlc-iso/tree/v2.0}}

\title{Hereditary permutations for simple types revisited\\[1ex]
\large A baby case for University Algebra}
\author{Thorsten Altenkirch\\University of Nottingham}
\date{\today}

\begin{document}
\maketitle

\begin{abstract}
  It is a classical result that the definable isomorphisms of the simply typed
  $\lambda$-calculus with function types are exactly the hereditary permutations,
  and that they are generated by $A\to B\to C\iso B\to A\to C$ and congruence. We
  give a new proof of this result that stays inside $\beta$-normal $\eta$-long
  typed terms and uses no completeness theorem: an induction that descends through
  both components of an isomorphism at once. It proceeds by showing that both
  components are linear terms: relevance follows from one of the inverse
  equations, and no copying from relevance of the other component. We then go one level up and prove a
  coherence theorem: two derivations of an isomorphism denote the same
  isomorphism if and only if they are equal modulo an explicit list of equations
  (groupoid laws, functoriality, involution, naturality and the braid equation),
  so these equations present the groupoid of isomorphisms, a hereditary version
  of the Coxeter presentation of the symmetric groups. All results are formalised
  in Agda, on intrinsically typed syntax with hereditary substitution, without
  postulates. The work is a case study for the syntactic completeness conjecture
  of University Algebra, a proposed solution of Tarski's high school algebra
  problem using dependent types, whose semantic counterpart is open. We explain
  the motivation and which parts of the method we expect to carry over.
\end{abstract}

\section{Introduction}\label{sec:intro}

\paragraph{Motivation.} University Algebra~\cite{alt08web,alt-ua} is a proposed answer
to Tarski's high school algebra problem: the usual laws for $1,+,\times$ and
exponentiation are incomplete for the positive integers~\cite{wilkie}, and so is
their categorified version, isomorphisms of types with sums~\cite{fdb}. With
dependent types ($\one$, $\two$, $\Sigma$, $\Pi$) a short list of basic
isomorphisms, including the type-theoretic axiom of choice, derives all high
school laws and Wilkie's identity, and it is conjectured to be complete for finite
sets. The conjecture splits into a semantic half (finite sets are complete for
definable isomorphism), which is open, and a \emph{syntactic} half: every
definable isomorphism is derivable, and in the strong, \emph{coherence}, form,
derivations modulo a chosen set of equations between derivations (the
\emph{level-2} equations) are in bijection with definable isomorphisms. Since the
semantic half is not available, a proof of the syntactic half cannot go through a
completeness theorem, unlike the classical proofs for simple types: it has to
analyse the normal forms of invertible terms directly and read off a derivation.

\paragraph{This paper.} We treat the baby case of this programme: the simply
typed $\lambda$-calculus with $\to$ only. Here the characterisation of definable
isomorphisms is classical (Section~\ref{sec:related}): the invertible
terms are the hereditary permutations~\cite{dezani}, and $\swap:A\to B\to C\iso
B\to A\to C$ with congruence generates them~\cite{bl,bdl}. Our aims are to find out
which lemmas a purely syntactic proof needs, to prove the coherence form, and to
check everything formally. The contributions are:
\begin{enumerate}
\item A direct proof that every invertible $\beta$-normal $\eta$-long term is a
  hereditary permutation (Theorem~\ref{thm:char}), by a \emph{descent} that analyses
  both components of an isomorphism simultaneously
  (Section~\ref{sec:char}). It uses no completeness theorem and does not pass
  through the untyped calculus. On the way it shows that both components of an
  isomorphism are \emph{linear}: relevance ($\lambda I$) follows from one inverse
  equation, and no copying from relevance of the other component. Linearity then
  makes the remaining step, reading off the permutation, immediate.
\item A coherence theorem (Theorem~\ref{thm:coh}): derivations modulo an explicit
  list of level-2 equations are in bijection with hereditary permutations
  (Section~\ref{sec:coh}). The proof is a hereditary version of the normal-form
  proof for the Coxeter presentation of the symmetric groups.
\item A formalisation of all results in Agda without postulates, on intrinsically
  typed de Bruijn syntax with hereditary substitution~\cite{ka}
  (Section~\ref{sec:formal}).
\item A discussion of which parts of the method should carry over to University
  Algebra (Section~\ref{sec:ua}).
\end{enumerate}

\section{Setting}\label{sec:setting}

\paragraph{Types.} $A::=X\mid A\to B$, with $X$ ranging over base types. Every
type can be written uniquely as $A_1\to\cdots\to A_n\to X$, which we write
$\vec A\Rightarrow X$.

\paragraph{Contexts and names.} The syntax is nameless, as in the
formalisation: contexts are lists of types, built from the empty context
$\varepsilon$ by extension, and variables are de~Bruijn indices, so all freshness
conditions become typing. To improve readability we nevertheless use names,
writing $\Gamma\triangleright x{:}A$ for the extension and $x$ for the
corresponding variable; every name refers to a position in its context, so the
text translates directly into de~Bruijn syntax.

\paragraph{Terms and normal forms.} Terms are those of the simply typed
$\lambda$-calculus, taken modulo $\beta\eta$. Every term has a unique
$\beta$-normal, $\eta$-long form:\footnote{The notation follows bidirectional
type checking~\cite{dk}: $T\ni t$, with the type on the left, says that $t$ is
\emph{checked} against the given type $T$, as normal forms are; $n\in T$, with the
type on the right, says that the type of $n$ is \emph{synthesised}, as it is for a
neutral term, from its head variable.}
\[
\Nf(\Gamma,\vec A\Rightarrow X)\ni\lambda\vec y.\,n,\qquad
n\in\Ne(\Gamma,\vec y{:}\vec A;X),\qquad
h\,t_1\ldots t_k\in\Ne(\Delta;X),
\]
where $h{:}\vec B\Rightarrow X$ is in $\Delta$ and $t_i\in\Nf(\Delta,B_i)$.

\paragraph{Hereditary substitution.} For $s\in\Nf(\Gamma,A)$ and
$t\in\Nf(\Gamma\triangleright x{:}A,C)$, the \emph{hereditary
substitution}~\cite{watkins,ka} $t[x=s]\in\Nf(\Gamma,C)$ substitutes and
normalises at once:
\[
(\lambda y.\,t)[x=s]=\lambda y.\,t[x=s],\qquad
(y\,\vec r)[x=s]=y\,(\vec r[x=s])\ \ (y\ne x),\qquad
(x\,\vec r)[x=s]=s\mathbin{@}(\vec r[x=s]),
\]
where $\vec r[x=s]$ substitutes in each argument, and \emph{hereditary
application} $s\mathbin{@}\vec u$ feeds the arguments to $s$ one at a time:
\[
s\mathbin{@}\varepsilon=s,\qquad
(\lambda z.\,t')\mathbin{@}(u,\vec u)=t'[z=u]\mathbin{@}\vec u .
\]
The recursion terminates because in the inner call $t'[z=u]$ the type of $z$ is
an argument type of $A$, hence smaller: it is by recursion on the type of the
substituted variable, and then on the term. Two features are used below. In the
last clause the head of $(x\,\vec r)[x=s]$ is the head of $s$, unless that head is
bound in $s$ (Lemma~\ref{lem:heads}). And if $z$ does not occur in $t'$, the
argument $u$ is erased, which is why counting occurrences needs $\lambda I$-terms
(Lemma~\ref{lem:count}).

\paragraph{$\eta$-expansion.} For a variable
$x{:}\vec A\Rightarrow X$, its $\eta$-long form is
\[
\eta x=\lambda y_1\ldots y_n.\,x\,(\eta y_1)\ldots(\eta y_n),
\]
defined by recursion on the type. It is the normal form of the term $x$, so
$\eta a\in\Nf(a{:}A,A)$ is the identity map on $A$. Composition of maps
$t\in\Nf(a{:}A,B)$ and $u\in\Nf(b{:}B,C)$ is $u\circ t=u[b=t]$; it is associative
and $\eta$ is its unit, by the substitution lemmas for hereditary substitution.

\paragraph{Relation to $\beta\eta$.} Hereditary substitution computes normal
forms: $t[x=s]$ is the $\beta\eta$-normal form of the ordinary substitution of $s$
for $x$ in $t$~\cite{watkins,ka}. More generally, normalisation of arbitrary terms
is defined by $\mathsf{nf}(x)=\eta x$, $\mathsf{nf}(\lambda x.\,t)=\lambda
x.\,\mathsf{nf}(t)$ and $\mathsf{nf}(t\,u)=t'[z=\mathsf{nf}(u)]$, where
$\mathsf{nf}(t)=\lambda z.\,t'$; this form exists because $\mathsf{nf}(t)$ has a
function type and is $\eta$-long. Moreover $\mathsf{nf}(t)=\mathsf{nf}(u)$ if and only if $t=_{\beta\eta}u$~\cite{ka}. So
$\Nf(\Gamma,A)$ is the set of terms of type $A$ in $\Gamma$ modulo $\beta\eta$, and
statements about normal forms with hereditary substitution, such as the
isomorphism equations below, are statements about terms modulo $\beta\eta$. The
formalisation works with normal forms only; this standard fact is the bridge to
the usual formulation.

\paragraph{Isomorphisms.} An \emph{isomorphism} $A\iso B$ is a pair
$f\in\Nf(a{:}A,B)$, $g\in\Nf(b{:}B,A)$ with
\[
g[b=f]=\eta a,\qquad f[a=g]=\eta b,
\]
as syntactic equalities of normal forms: both composites are the identity. We
call $f$ \emph{invertible}. The inverse is unique: if $g'$ is another, then
$g'=g'\circ(f\circ g)=(g'\circ f)\circ g=g$. So being invertible is a property of
$f$ alone.

\paragraph{Derivations.} The relation $A\iso_{\mathrm{ax}}B$ is generated by the axiom
\[
\swap:\;A\to B\to C\;\iso\;B\to A\to C
\]
together with $\mathrm{id}$, ${}^{-1}$, composition $;$ (in diagrammatic order) and
the congruence $\varphi\to\psi:A\to B\iso A'\to B'$ for $\varphi:A\iso A'$,
$\psi:B\iso B'$. \emph{Derivations} are the terms of this signature. Each
derivation $\varphi$ has a \emph{meaning} $\sem\varphi=(f,g)$, an isomorphism,
defined by recursion on $\varphi$. Both parts of $\sem\swap$ are $\lambda b\,a.\,u\,a\,b$,
with input variable $u$; $\mathrm{id}$, ${}^{-1}$ and $;$ are interpreted by the
identity, by exchanging $f$ and $g$, and by composition; and the first part of
$\sem{\varphi\to\psi}$ is $\lambda a'.\,f_\psi\,(u\,(g_\varphi\,a'))$. We also write
$\sem\varphi$ for its first part $f$.

\paragraph{Level-2 equations.} The equations between derivations are:
\begin{itemize}
\item the groupoid laws, and congruence for $;$, ${}^{-1}$ and $\to$;
\item functoriality of $\to$: $\mathrm{id}\to\mathrm{id}=\mathrm{id}$ and
  $(\varphi;\varphi')\to(\psi;\psi')=(\varphi\to\psi);(\varphi'\to\psi')$;
\item $\swap;\swap=\mathrm{id}$;
\item naturality of $\swap$:
  $\swap;(\psi\to\varphi\to\chi)=(\varphi\to\psi\to\chi);\swap$;
\item the braid (Yang--Baxter) equation on $A\to B\to C\to D$:
  $\swap;(\mathrm{id}\to\swap);\swap=(\mathrm{id}\to\swap);\swap;(\mathrm{id}\to\swap)$.
\end{itemize}
We write $\varphi\sim\psi$ for equality modulo these equations. Swaps at
non-adjacent positions commute: this is naturality with $\chi=\swap$. Together
these are the Coxeter relations of the symmetric groups, applied hereditarily.

\paragraph{Normal forms of types.} Modulo $\swap$, a type $\vec A\Rightarrow X$ is
determined by $X$ and the \emph{multiset} $\mset{A_1,\ldots,A_n}$. Hereditarily,
types modulo $\iso_{\mathrm{ax}}$ are rooted unordered trees: the root is labelled
by $X$, and each subtree is one $A_i$.

\section{Hereditary permutations and the main results}\label{sec:results}

\begin{definition}
  The \emph{hereditary permutations} $\HP(A,B)\subseteq\Nf(a{:}A,B)$, for
  $A=\vec A\Rightarrow X$ and $B=\vec B\Rightarrow X$ with $|\vec A|=|\vec B|=n$,
  are the normal terms
  \[
  \lambda y_1\ldots y_n.\;a\;\big(h_1[z_1=y_{\pi 1}]\big)\ldots
  \big(h_n[z_n=y_{\pi n}]\big),
  \]
  where $\pi$ is a permutation of $\{1,\ldots,n\}$ and
  $h_i\in\HP(B_{\pi i},A_i)$ has free variable $z_i$. The case $n=0$ gives
  $\eta a$ at a base type.
\end{definition}

\begin{proposition}[Easy direction]\label{prop:easy}
  \leavevmode
  \begin{enumerate}
  \item Every hereditary permutation is invertible, and the inverse is again a
    hereditary permutation (with $\pi^{-1}$ and the inverses of the $h_i$).
  \item The meaning of every derivation is a hereditary permutation: $\swap$
    gives a transposition, and $\to$-congruence gives the $h_i$.
  \item Conversely, every hereditary permutation $h$ is the meaning of a
    derivation $\mathsf{der}\,h$, which decomposes $\pi$ into adjacent
    transpositions (Section~\ref{sec:coh}).
  \end{enumerate}
\end{proposition}

\begin{theorem}[Characterisation]\label{thm:char}
  Every invertible normal term is a hereditary permutation.
\end{theorem}

\begin{theorem}[Coherence]\label{thm:coh}
  Taking meanings, $\varphi\mapsto\sem\varphi$, is a bijection from derivations of
  $A\iso_{\mathrm{ax}}B$ modulo $\sim$ to $\HP(A,B)$. In particular,
  $\varphi\sim\psi$ if and only if $\sem\varphi=\sem\psi$.
\end{theorem}

Theorem~\ref{thm:char} together with Proposition~\ref{prop:easy} gives
completeness: $A\iso B$ iff $A\iso_{\mathrm{ax}}B$ iff $A$ and $B$ have the same
tree. Theorem~\ref{thm:coh} is the analogue, for $\to$, of the coherence form of
syntactic completeness for University Algebra.

\section{Proof of the characterisation}\label{sec:char}

We give the proof with context morphisms, i.e.\ simultaneous substitutions between
contexts.

Throughout, $f\in\Nf(a{:}A,B)$ and $g\in\Nf(b{:}B,A)$ are mutually inverse, with
$A=\vec A\Rightarrow X$ and $B=\vec B\Rightarrow Y$; the equations are
$f[a=g]=\eta b$ and $g[b=f]=\eta a$.

\paragraph{Strategy.} The two equations are used one at a time. Each lemma is
stated for \emph{one} equation and then applied to both, using the symmetry
$(f,g)\mapsto(g,f)$. The two equations interact at two points: the matching of
heads in Lemma~\ref{lem:descent}(c), and no copying (Lemma~\ref{lem:linmp}),
which uses relevance of the other component. The proof goes
\[
\text{descent}\;\Rightarrow\;\text{relevance}\;\Rightarrow\;\text{no copying}
\;\Rightarrow\;\text{hereditary permutation},
\]
and on the way shows that $f$ and $g$ are linear terms
(Corollary~\ref{cor:lin}).

\subsection{Substitutions}

Contexts are lists of types, and a \emph{telescope} is a context that extends
another one: $\Gamma\mathbin{+\!\!+}\Omega$. We use $\Gamma,\Delta$ for contexts
and the other Greek capitals $\Theta,\Xi,\Phi,\Psi,\Omega$ for telescopes. A
\emph{substitution}
$\rho\in\Sub(\Gamma,\Delta)$ is a normal form in $\Gamma$ for each type of
$\Delta$. We write:
\begin{itemize}
\item $\rho_i$ for the $i$th component of $\rho$;
\item $t[\rho]\in\Nf(\Gamma,C)$ for simultaneous hereditary substitution, where
  $t\in\Nf(\Delta,C)$; it is defined like $t[x=s]$, with
  $(v\,\vec r)[\rho]=\rho_v\mathbin{@}(\vec r[\rho])$ for every variable $v$ of
  $\Delta$ (under binders, $\rho$ is weakened and extended by the bound variables);
\item $\delta\circ\rho$ for composition, applying $t\mapsto t[\rho]$ to each component
  of $\delta$, so $(\delta\circ\rho)_i=\delta_i[\rho]$;
\item $\langle\rho,\delta\rangle\in\Sub(\Gamma,\Delta\mathbin{+\!\!+}\Omega)$ for
  pairing, where $\rho\in\Sub(\Gamma,\Delta)$ and $\delta\in\Sub(\Gamma,\Omega)$;
\item $\mathsf p\in\Sub(\Gamma\mathbin{+\!\!+}\Omega,\Gamma)$ for the weakening
  projection, and $\mathsf q_\Omega\in\Sub(\Gamma\mathbin{+\!\!+}\Omega,\Omega)$
  for the $\eta$-expanded variables of $\Omega$.
\end{itemize}
Contexts and substitutions form a category, and $\langle-,-\rangle$, $\mathsf p$,
$\mathsf q$ satisfy the usual laws. In particular, $\rho\circ\mathsf p$ is $\rho$
weakened past $\Omega$, and $\langle\mathsf p,\mathsf q_\Omega\rangle$ is the
identity on $\Gamma\mathbin{+\!\!+}\Omega$. A term $\lambda\Omega.\,h\,\vec n$ is
\emph{head-free} if its head $h$ is not bound in $\Omega$. Components of $\mathsf
p$ and $\mathsf q$ are head-free.

\begin{lemma}[No creation]\label{lem:noncreation}
  Let $t\in\Nf(\Delta,C)$ and $\rho\in\Sub(\Gamma,\Delta)$. If a variable $z$ of
  $\Gamma$ occurs in $t[\rho]$, then $z$ occurs in $\rho_v$ for some variable $v$
  of $\Delta$ that occurs in $t$. If $t=\lambda\Omega.\,M$, then
  $t[\rho]=\lambda\Omega.\,M[\langle\rho\circ\mathsf p,\mathsf q_\Omega\rangle]$.
\end{lemma}

\begin{lemma}[Heads]\label{lem:heads}
  Let $t=\lambda\Omega.\,h\,\vec r\in\Nf(\Delta,C)$ and
  $\rho\in\Sub(\Gamma,\Delta)$.
  \begin{itemize}
  \item If $h$ is bound in $\Omega$, the head of $t[\rho]$ is the same variable of
    $\Omega$.
  \item If $h=v$ is a variable of $\Delta$ and $\rho_v$ is head-free with head $h'$,
    then the head of $t[\rho]$ is $h'$, weakened past $\Omega$.
  \end{itemize}
\end{lemma}

Both lemmas are by induction on $t$, and on the type for the second clause of
Lemma~\ref{lem:heads}.

\subsection{The descent}

\paragraph{Idea.} We want to read off the structure of $f$ top-down: first its
head, then the heads of its arguments, and so on. The only information is the pair
of equations $g[b=f]=\eta a$ and $f[a=g]=\eta b$, and they are about the whole terms.
Write $f=\lambda\vec y.\,\mathsf a\,\vec t$ and $g=\lambda\vec x.\,\mathsf b\,\vec s$.
In $f[a=g]$ the head $\mathsf a$ is replaced by $g$, so the arguments $\vec t$
are substituted into the body of $g$. Comparing with $\eta b$ says that the
head of $s_j$ must be the variable that receives $t_{\sigma j}$, and that the
arguments of these heads again satisfy two equations of the same kind, one level
deeper. The difference is that the deeper equations live in a larger context:
the variables bound on the way down, and the parts of $f$ and $g$ that were already
substituted.

A \emph{matched pair} is the general form of such a pair of equations. $P$ is a
family of subterms of $f$ at the current depth, $Q$ is the matching family of
subterms of $g$, and the parameter substitutions $\alpha_1,\beta_1$ record what has
been substituted so far. The initial pair is $(f,g)$ itself. Lemma~\ref{lem:descent}
shows that the form is preserved when we step from a pair of heads to a pair of
their arguments. So the \emph{descent} is an induction over the normal forms $f$
and $g$, but over both at once, along matched positions: argument $i$ of a head in
$f$ goes with argument $\tau i$ in $g$. It is not structural induction on one term,
because each step uses one equation for the heads of $P$ and the other for the heads
of $Q$ (part (c)). It terminates because the pair of types gets smaller.

\begin{definition}\label{def:matched}
  A \emph{matched pair} consists of:
  \begin{itemize}
  \item contexts $\Gamma_1,\Gamma_2$ (the \emph{bases}) and $\Gamma_P,\Gamma_Q$
    (the \emph{parameters});
  \item telescopes $\Theta$ and $\Xi$ (the \emph{targets});
  \item $P\in\Sub(\Gamma_P\mathbin{+\!\!+}\Xi,\Theta)$ and
    $Q\in\Sub(\Gamma_Q\mathbin{+\!\!+}\Theta,\Xi)$;
  \item parameter substitutions $\alpha_1\in\Sub(\Gamma_1,\Gamma_P)$,
    $\beta_1\in\Sub(\Gamma_1,\Gamma_Q)$, $\alpha_2\in\Sub(\Gamma_2,\Gamma_P)$ and
    $\beta_2\in\Sub(\Gamma_2,\Gamma_Q)$, all of whose components are head-free;
  \end{itemize}
  such that
  \begin{align*}
  (P)\quad & P\circ\big\langle\alpha_1\circ\mathsf p,\;Q\circ\langle\beta_1\circ\mathsf p,\mathsf q_\Theta\rangle\big\rangle
       =\mathsf q_\Theta\quad\text{in }\Sub(\Gamma_1\mathbin{+\!\!+}\Theta,\Theta),\\
  (Q)\quad & Q\circ\big\langle\beta_2\circ\mathsf p,\;P\circ\langle\alpha_2\circ\mathsf p,\mathsf q_\Xi\rangle\big\rangle
       =\mathsf q_\Xi\quad\text{in }\Sub(\Gamma_2\mathbin{+\!\!+}\Xi,\Xi).
  \end{align*}
\end{definition}

Componentwise, $(P)$ says that
$P_i[\langle\alpha_1\circ\mathsf p,Q\circ\langle\beta_1\circ\mathsf p,\mathsf
q_\Theta\rangle\rangle]=\eta\,x_i$ for each variable $x_i$ of $\Theta$. Here
$\langle\beta_1\circ\mathsf p,\mathsf q_\Theta\rangle$ instantiates the parameters of $Q$
by $\beta_1$ and keeps the targets $\Theta$, and $P$ is then applied to the result,
with its own parameters instantiated by $\alpha_1$. So $(P)$ says that $Q$
followed by $P$ is the identity on $\Theta$, relative to the parameters, and $(Q)$
says the same with $P$ and $Q$ swapped.

The definition is symmetric under
$(\Gamma_1,\Gamma_P,\Theta,P,\alpha_1,\alpha_2)\leftrightarrow(\Gamma_2,\Gamma_Q,\Xi,Q,\beta_2,\beta_1)$,
which exchanges $(P)$ and $(Q)$. There are no freshness conditions: $\alpha_1$ lives
over $\Gamma_1$, so it cannot mention the targets $\Theta$.

\paragraph{The initial pair.} All bases and parameters are empty,
$\Theta=\varepsilon\triangleright b{:}B$, $\Xi=\varepsilon\triangleright a{:}A$,
$P=(f)$, $Q=(g)$, and the parameter substitutions are empty. Then $(P)$ is
$f[a=g]=\eta b$ and $(Q)$ is $g[b=f]=\eta a$.

\begin{lemma}[Descent]\label{lem:descent}
  For a matched pair:
  \begin{enumerate}
  \item[(a)] (one-sided) the head of $P_i=\lambda \Phi_i.\,h_i\,\vec t$ is a variable
    $y_{\tau i}$ of $\Xi$, and dually the head of $Q_j=\lambda \Psi_j.\,k_j\,\vec s$ is a
    variable $x_{\sigma j}$ of $\Theta$. So $\tau:|\Theta|\to|\Xi|$ and $\sigma:|\Xi|\to|\Theta|$ are
    maps between positions.
  \item[(b)] (one-sided) every variable of $\Phi_i$ occurs in the body of $P_i$; dually
    for $\Psi_j$.
  \item[(c)] $\sigma$ and $\tau$ are mutually inverse; in particular $|\Theta|=|\Xi|$.
  \item[(d)] for each $i$, let $j=\tau i$, $\Phi=\Phi_i$ and $\Psi=\Psi_j$. Then the argument
    families $\vec s\in\Sub(\Gamma_Q\mathbin{+\!\!+}\Theta\mathbin{+\!\!+}\Psi,\Phi)$ of $Q_j$
    and $\vec t\in\Sub(\Gamma_P\mathbin{+\!\!+}\Xi\mathbin{+\!\!+}\Phi,\Psi)$ of $P_i$ form
    a matched pair with
    \begin{gather*}
    P'=\vec s,\quad Q'=\vec t,\quad \Theta'=\Phi,\quad \Xi'=\Psi,\quad
    \Gamma_P'=\Gamma_Q\mathbin{+\!\!+}\Theta,\quad \Gamma_Q'=\Gamma_P\mathbin{+\!\!+}\Xi,\\
    \Gamma_1'=\Gamma_1\mathbin{+\!\!+}\Theta,\qquad \Gamma_2'=\Gamma_2\mathbin{+\!\!+}\Xi,\\
    \alpha'_1=\langle\beta_1\circ\mathsf p,\mathsf q_\Theta\rangle,\qquad
    \beta'_1=\langle\alpha_1\circ\mathsf p,\,Q\circ\langle\beta_1\circ\mathsf p,\mathsf q_\Theta\rangle\rangle,\\
    \alpha'_2=\langle\beta_2\circ\mathsf p,\,P\circ\langle\alpha_2\circ\mathsf p,\mathsf q_\Xi\rangle\rangle,\qquad
    \beta'_2=\langle\alpha_2\circ\mathsf p,\mathsf q_\Xi\rangle.
    \end{gather*}
    So the new parameter substitution $\beta'_1$ is the substitution of the old
    equation $(P)$, and $\alpha'_2$ is that of $(Q)$.
  \end{enumerate}
\end{lemma}

\begin{proof}
  (a) Apply Lemma~\ref{lem:heads} to $(P)$ at component $i$. The right-hand side is
  $\eta$ of the $i$th variable of $\Theta$, whose head is that variable.
  \begin{itemize}
  \item If $h_i$ is bound in $\Phi_i$, the head of the left-hand side is in $\Phi_i$.
  \item If $h_i$ is in $\Gamma_P$, the head of the left-hand side is the head of
    a component of $\alpha_1\circ\mathsf p$, which is in $\Gamma_1$, since $\alpha_1$ is
    head-free.
  \end{itemize}
  Neither is in $\Theta$, so $h_i$ is in $\Xi$. The dual is the same argument applied to
  $(Q)$.

  (b) By Lemma~\ref{lem:noncreation}, the left-hand side of $(P)$ at $i$ is
  $\lambda \Phi_i.\,\ldots$ with the same binders as $\eta\,x_i$, and each of them
  occurs in $\eta\,x_i$. No component of a substitution mentions $\Phi_i$, so by no
  creation each variable of $\Phi_i$ occurs in the body of $P_i$.

  (c), (d) In $(P)$ at $i$, the head $h_i$ is replaced by
  $Q_{\tau i}[\langle\beta_1\circ\mathsf p,\mathsf q_\Theta\rangle]$. By (a) for $Q$, its head
  is the $\sigma\tau i$th variable of $\Theta$, which $\langle\beta_1\circ\mathsf p,\mathsf
  q_\Theta\rangle$ maps to an $\eta$-expanded variable. So the left-hand side is
  $\lambda \Phi.\,x_{\sigma\tau i}\,(\ldots)$. Comparing heads with $\eta\,x_i$ gives
  $\sigma\tau=\mathrm{id}$. Comparing arguments gives
  \[
  (P')\qquad \vec s\circ\big\langle\alpha'_1\circ\mathsf p,\;
     \vec t\circ\langle\beta'_1\circ\mathsf p,\mathsf q_\Phi\rangle\big\rangle=\mathsf q_\Phi .
  \]
  This uses the substitution lemma to regroup the iterated substitution as a
  single one. Dually, $(Q)$ at $j=\tau i$ gives $\tau\sigma=\mathrm{id}$ and
  equation $(Q')$ of the new pair. The new parameter substitutions are head-free:
  the components of $\mathsf q$ are $\eta$-expanded variables, and the components
  of $Q\circ\ldots$ and $P\circ\ldots$ have heads in $\Theta$ and $\Xi$ by (a).
\end{proof}

\begin{example}[Descent for the swap]\label{ex:descent}
  Let $X_1,X_2,X$ be base types, $A=X_1\to X_2\to X$ and $B=X_2\to X_1\to X$, and
  \[
  f=\lambda x_2x_1.\,a\,x_1\,x_2\in\Nf(a{:}A,B),\qquad
  g=\lambda y_1y_2.\,b\,y_2\,y_1\in\Nf(b{:}B,A),
  \]
  where $x_k,y_k{:}X_k$. In the initial pair, $\Theta=(b{:}B)$ and $\Xi=(a{:}A)$,
  and the heads of $P_1=f$ and $Q_1=g$ are $a$ and $b$, so $\sigma$ and $\tau$ are
  the identity on $\{1\}$. Part~(d), with $\Phi=(x_2{:}X_2,x_1{:}X_1)$ and
  $\Psi=(y_1{:}X_1,y_2{:}X_2)$, gives a new pair of size $n=2$:
  \begin{gather*}
  \Theta'=\Phi,\quad \Xi'=\Psi,\quad
  P'=(x_2=y_2,\;x_1=y_1)\in\Sub((b{:}B)\mathbin{+\!\!+}\Psi,\Phi),\\
  Q'=(y_1=x_1,\;y_2=x_2)\in\Sub((a{:}A)\mathbin{+\!\!+}\Phi,\Psi),\\
  \Gamma_1'=\Gamma_P'=(b{:}B),\quad \Gamma_2'=\Gamma_Q'=(a{:}A),\quad
  \alpha'_1=(\eta b),\quad \beta'_1=(g),\quad
  \alpha'_2=(f),\quad \beta'_2=(\eta a).
  \end{gather*}
  So $P'$ consists of the arguments of the head $b$ of $g$, and $Q'$ of the
  arguments of the head $a$ of $f$. The parameter substitutions record that $a$
  has been replaced by $g$ in $(P)$, and $b$ by $f$ in $(Q)$. Equation $(P')$ over
  $(b{:}B)\mathbin{+\!\!+}\Phi$ is checked as follows:
  \begin{itemize}
  \item $Q'\circ\langle\beta'_1\circ\mathsf p,\mathsf q_\Phi\rangle$ substitutes $g$ for
    $a$ in $Q'$, which gives $Q'$ again, since $a$ does not occur;
  \item $P'\circ\langle\alpha'_1\circ\mathsf p,Q'\rangle$ substitutes $b$ for $b$,
    $x_1$ for $y_1$ and $x_2$ for $y_2$, which gives $(x_2=x_2,\;x_1=x_1)$,
    and this is $\mathsf q_\Phi$.
  \end{itemize}
  $(Q')$ is checked in the same way. By name, each $x_k$ is matched with $y_k$,
  but $x_k$ is at position $3-k$ of $\Phi$ and $y_k$ at position $k$ of $\Psi$.
  So $\tau'$ and $\sigma'$ are the swap $1\leftrightarrow2$, the permutation of the
  isomorphism. Since $X_1$ and $X_2$ are base types, the components of $P'$ and
  $Q'$ have no binders and no arguments, and the descent stops.
\end{example}

\begin{corollary}[Relevance]\label{lem:rel}
  $f$ and $g$ are $\lambda I$-terms: every bound variable occurs in the body of its
  binder.
\end{corollary}
\begin{proof}
  In an $\eta$-long normal form, every $\lambda$ is a top binder of the whole term
  or of an argument of a head variable. Starting from the initial pair,
  Lemma~\ref{lem:descent}(d) reaches every argument of every head in $f$ and in
  $g$: $P_i=\lambda \Phi_i.\,h_i\,\vec t$ passes all of $\vec t$ on. Part~(b)
  applies at each level. The descent terminates: a step goes from the pair of
  types $(\Theta_i,\Xi_{\tau i})$ to a pair of argument types of $\Theta_i$ and of $\Xi_{\tau i}$,
  and this relation is well founded.
\end{proof}

\begin{remark}[One equation is not enough]\label{rem:retract}
  A single equation $g\circ f=\mathrm{id}$ makes $A$ a retract of $B$, and
  retractions of simple types are much less rigid~\cite{ru}:
  \begin{itemize}
  \item The head of a section need not be its input:
    $\Theta\lhd(\Theta\to \Theta)\to \Theta$ via $f=\lambda a.\,\lambda h.\,h\,a$ and
    $g=\lambda b.\,b\,(\lambda z.\,z)$.
  \item A section may erase: for $f=\lambda a.\,\lambda h\,x.\,a\,x$ and
    $g=\lambda b.\,\lambda x.\,b\,(\lambda z.\,z)\,x$, we have $g\circ f=\mathrm{id}$
    but $f$ is not $\lambda I$. In terms of the descent, $(Q)$ holds but $(P)$
    fails, and part~(b) for $(P)$ is exactly what detects the vacuous $\lambda h$.
  \item A section may copy: for $f=\lambda a.\,\lambda h.\,h\,a\,a$ and
    $g=\lambda b.\,b\,(\lambda x\,y.\,x)$, we have $g\circ f=\mathrm{id}$, but $a$
    occurs twice in $f$. Here $g$ is not $\lambda I$, and this is what
    Lemma~\ref{lem:linmp} below exploits: no copying in $f$ follows from
    relevance of $g$.
  \end{itemize}
  This is why the descent needs the heads of \emph{both} families at each level,
  taking one from each equation, and why no copying needs relevance of the other
  side. Relevance itself is one-sided.
\end{remark}

\subsection{Linearity}

Relevance says that every bound variable occurs \emph{at least} once. We now show
that it occurs \emph{at most} once, so that $f$ and $g$ are linear terms. Write
$\#_v(t)$ for the number of occurrences of the variable $v$ in $t$. Unlike
relevance, no copying uses both equations: the components of the substitution in
$(P)$ come from $Q$, and they are $\lambda I$ by part~(b) applied to $(Q)$.

\begin{lemma}[Counting]\label{lem:count}
  If $t$ and all components of $\rho$ are $\lambda I$, then so is $t[\rho]$, and
  for all variables $v$ and $z$,
  \[
  \#_v(t)\cdot\#_z(\rho_v)\;\le\;\#_z(t[\rho]).
  \]
\end{lemma}
\begin{proof}
  Hereditary substitution loses occurrences only through a $\beta$-step
  $(\lambda q.\,N)\,r$ in which $q$ does not occur in $N$, and $\lambda I$ rules
  this out. It may copy, so we only get an inequality. First consider a single
  substitution $t[x=a]$ and the application $\mathsf{app}\,t\,\vec r$, for
  $\lambda I$-terms. We prove simultaneously that the result is $\lambda I$ and
  that, for $z\ne x$,
  \[
  \#_z(t)+\#_x(t)\cdot\#_z(a)\le\#_z(t[x=a]),\qquad
  \#_z(t)+\textstyle\sum_i\#_z(r_i)\le\#_z(\mathsf{app}\,t\,\vec r).
  \]
  The induction is the one that defines hereditary substitution: on $t$ for
  substitution, on $\vec r$ and the type for application. The key case is
  $\mathsf{app}\,(\lambda q.\,N)\,(r,\vec r)$: by the first inequality
  $\#_z(N[q=r])\ge\#_z(N)+\#_q(N)\cdot\#_z(r)\ge\#_z(N)+\#_z(r)$, because
  $\#_q(N)\ge1$ by $\lambda I$. The simultaneous case follows by induction on $t$,
  lifting $\rho$ under binders. The lifted substitution is again $\lambda I$,
  because weakenings and $\eta$-expanded variables are, and the result is
  $\lambda I$ because a binder $q$ of $t$ satisfies
  $\#_q(t[\rho])\ge\#_q(t)\cdot\#_q(\eta q)=\#_q(t)\ge1$.
\end{proof}

Starting from the initial pair, the components $P,Q$ and the parameter
substitutions of every pair reached by the descent are $\lambda I$: $P$ and $Q$
are families of subterms of $f$ and $g$, which are $\lambda I$ by
Corollary~\ref{lem:rel}, and the parameters are built from them by substitution,
which preserves $\lambda I$ by Lemma~\ref{lem:count}. Call such a pair a
\emph{$\lambda I$ pair}.

\begin{lemma}[No copying]\label{lem:linmp}
  In a $\lambda I$ matched pair, let $P_i=\lambda \Phi_i.\,y_{\tau i}\,\vec t$ as in
  Lemma~\ref{lem:descent}(a). Then:
  \begin{enumerate}
  \item[(i)] no variable of $\Gamma_P$ occurs in $P_i$;
  \item[(ii)] a variable $y_j$ of $\Xi$ occurs in $P_i$ only if $j=\tau i$, and
    $y_{\tau i}$ occurs in $P_i$ exactly once.
  \end{enumerate}
  Dually for $Q$.
\end{lemma}
\begin{proof}
  Let $\rho=\langle\alpha_1\circ\mathsf p,Q\circ\langle\beta_1\circ\mathsf p,\mathsf
  q_\Theta\rangle\rangle$ be the substitution of $(P)$; its components are
  $\lambda I$. By Lemma~\ref{lem:count} applied to $(P)$ at component $i$, for
  every variable $v$ of $\Gamma_P\mathbin{+\!\!+}\Xi$ and every $z$,
  \[
  \#_v(P_i)\cdot\#_z(\rho_v)\le\#_z(\eta\,x_i),
  \]
  and the right-hand side is $1$ if $z=x_i$ and $0$ otherwise. Choose for $z$ the
  head of $\rho_v$, so that $\#_z(\rho_v)\ge1$.
  \begin{itemize}
  \item If $v$ is in $\Gamma_P$, then $\rho_v$ is a component of
    $\alpha_1\circ\mathsf p$, whose head is in $\Gamma_1$ because $\alpha_1$ is
    head-free. So $z\ne x_i$ and $\#_v(P_i)=0$.
  \item If $v=y_j$, then $\rho_v=Q_j[\ldots]$ has head $z=x_{\sigma j}$ by
    Lemma~\ref{lem:descent}(a). So $\#_{y_j}(P_i)=0$ unless $\sigma j=i$, i.e.\
    $j=\tau i$, and $\#_{y_{\tau i}}(P_i)\le1$. Since $y_{\tau i}$ is the head of
    $P_i$, it occurs at least once.\qedhere
  \end{itemize}
\end{proof}

\begin{corollary}[Linearity]\label{cor:lin}
  $f$ and $g$ are linear: the input occurs exactly once, and every bound variable
  occurs exactly once in the body of its binder.
\end{corollary}
\begin{proof}
  For the input, apply Lemma~\ref{lem:linmp}(ii) to the initial pair, where
  $\Xi=(a)$ and $P_1=f$. A bound variable $w$ of $f$ is a binder of some
  $P_i=\lambda \Phi_i.\,y\,\vec t$ in a pair reached by the descent. By
  Lemma~\ref{lem:descent}(d), $\vec t$ is the family $Q'$ of the next pair, whose
  targets are $\Theta'=\Phi_i$. By parts (a) and (c) for the next pair, $w$ is the head
  of exactly one component $t_k$, and by Lemma~\ref{lem:linmp}(ii) for $Q'$, it
  occurs in $t_k$ exactly once and in no other component. It is not the head $y$
  of $P_i$, so it occurs once in the body of $P_i$. Every binder is reached, as
  in the proof of Corollary~\ref{lem:rel}. Dually for $g$.
\end{proof}

For the characterisation only one level of Lemma~\ref{lem:linmp} is used.

\begin{lemma}[No stray variables]\label{lem:stray}
  Write $f=\lambda\vec B.\,\mathsf{a}\,\vec t$ and
  $g=\lambda\vec A.\,\mathsf{b}\,\vec s$, where $\mathsf a$ and $\mathsf b$ are the
  input variables. Then $t_i$ is a weakening of a term in the single variable at
  position $\tau i$ of $\vec B$, and dually for $s_j$.
\end{lemma}
\begin{proof}
  The arguments $\vec t$ form the family $Q$ of the pair one level below the
  initial pair, with parameters $\Gamma_Q=(\mathsf a)$ and targets $\Theta=\vec B$. By
  Lemma~\ref{lem:linmp}, $\mathsf a$ does not occur in $t_i$, and the only target
  variable occurring in $t_i$ is its head, the variable at position $\tau i$.
  Dually for $s_j$.
\end{proof}

\begin{proof}[Proof of Theorem~\ref{thm:char}]
  We prove that $f$ and $g$ are both hereditary permutations, by well-founded
  induction on the pair of types $(A,B)$. They are treated together because the
  recursive call reverses the direction (step~4). In names, write
  $A=A_1\ldots A_m\Rightarrow X$, $B=B_1\ldots B_n\Rightarrow Y$ and
  $f=\lambda y_1\ldots y_n.\,h\,t_1\ldots t_k$, with $y_j:B_j$, and
  $g=\lambda x_1\ldots x_m.\,b\,s_1\ldots s_n$, with $x_i:A_i$ (in the
  formalisation the $y_j$ are the telescope $\vec B$ and $h$ is a de~Bruijn
  index).
  \begin{enumerate}
  \item \emph{Head.} By Lemma~\ref{lem:descent}(a) for the initial pair, $h$ is
    the input variable $a$. So $k=m$, and $X=Y$, because $a\,t_1\ldots t_m$ has
    type $X$ and is the body of a term of type $B$.
  \item \emph{Permutation.} The descent assigns to the argument $t_i$ the
    position $\tau i$ in $\vec B$, and to the argument $s_j$ of $g$ the position
    $\sigma j$ in $\vec A$. By Lemma~\ref{lem:descent}(c) these are mutually
    inverse, so $m=n$ and $\tau$ is a permutation of $\{1,\ldots,n\}$.
  \item \emph{Arguments.} By Lemma~\ref{lem:stray}, $t_i$ mentions only
    $y_{\tau i}$. Strengthening gives
    $t'_i\in\Nf(z_i{:}B_{\tau i},A_i)$ with $t_i=t'_i[z_i=y_{\tau i}]$.
    Likewise $s_{\tau i}=s'_{\tau i}[w_i=x_i]$ with
    $s'_{\tau i}\in\Nf(w_i{:}A_i,B_{\tau i})$.
  \item \emph{Inverse at a smaller type.} The first-level equations of the descent
    become $t'_i[z_i=s'_{\tau i}]=\eta\,w_i$ and $s'_{\tau i}[w_i=t'_i]=\eta\,z_i$.
    So $t'_i$ is an isomorphism $B_{\tau i}\iso A_i$, with inverse
    $s'_{\tau i}$, and the pair $(B_{\tau i},A_i)$ is smaller than $(A,B)$. By
    induction, $t'_i\in\HP(B_{\tau i},A_i)$.
  \item \emph{Assembly.} Putting the steps together,
    \[
    f=\lambda y_1\ldots y_n.\;a\;\big(t'_1[z_1=y_{\tau 1}]\big)\ldots
      \big(t'_n[z_n=y_{\tau n}]\big),
    \]
    which is the form of the definition of $\HP(A,B)$ with $\pi=\tau$ and
    $h_i=t'_i$. The base case $n=0$ is $f=\eta\,a$.
  \end{enumerate}
  By the symmetry $(f,g)\mapsto(g,f)$, $g$ is the hereditary permutation with
  $\pi=\sigma=\tau^{-1}$ and components $s'_j$. In the formalisation, this
  assembly is \texttt{characterisation-HP} in \texttt{STLC.Link}.
\end{proof}

\begin{remark}[Retracts in both directions]
  Suppose $g\circ f=\mathrm{id}_A$, and $B$ is a retract of $A$ by \emph{some}
  other pair of terms. In every interpretation of the base types by non-empty
  finite sets, $f$ is then injective and $|B|\le|A|$, so $f$ is a bijection and
  $f\circ g=\mathrm{id}$ holds semantically. By Statman's finite completeness
  theorem~\cite{statman,sd}, $f\circ g=_{\beta\eta}\mathrm{id}$. So retracts in
  both directions give an isomorphism. We do not use this: the syntactic proof needs
  both equations anyway, and the semantic argument is exactly the kind of
  completeness theorem that is not available for University Algebra.
\end{remark}

\section{Proof of coherence}\label{sec:coh}

In this section we prove Theorem~\ref{thm:coh}: two derivations are equal modulo
the level-2 equations if and only if they have the same meaning, and every
hereditary permutation is the meaning of a derivation.

\paragraph{Plan.} We choose for every hereditary permutation $h\in\HP(A,B)$ a
\emph{canonical derivation} $\mathsf{der}\,h:A\iso_{\mathrm{ax}}B$, defined below.
The theorem then follows from three facts:
\begin{enumerate}
\item[(S)] \emph{soundness}: if $\varphi\sim\psi$ then $\sem\varphi=\sem\psi$;
\item[(N)] \emph{normal forms}: every derivation $\varphi:A\iso_{\mathrm{ax}}B$
  satisfies $\varphi\sim\mathsf{der}\,h$ for some $h\in\HP(A,B)$;
\item[(St)] \emph{stability}: $\sem{\mathsf{der}\,h}=h$.
\end{enumerate}
Indeed, $\sem{-}$ is well defined on derivations modulo $\sim$ by (S), surjective
by (St), and injective: if $\sem\varphi=\sem\psi$, write
$\varphi\sim\mathsf{der}\,h$ and $\psi\sim\mathsf{der}\,h'$ by (N); then
$h=\sem{\mathsf{der}\,h}=\sem\varphi=\sem\psi=\sem{\mathsf{der}\,h'}=h'$ by (St)
and (S), so $\varphi\sim\psi$.

In the terminology of normalisation by evaluation, $\sem{-}$ is the normaliser,
hereditary permutations are the normal forms and $\mathsf{der}$ is their
embedding. (S) is soundness, (N) gives completeness,
$\varphi\sim\mathsf{der}\,\sem\varphi$, and (St) is stability. As in NbE,
stability is an equation between normal forms, not one up to $\sim$.

The three facts are of different natures. (N) is pure combinatorics on
derivations and uses no substitution at all; it is a hereditary version of the
normal-form proof for the Coxeter presentation of the symmetric groups, and it is
proved in Section~\ref{sec:coh-nf}. (S) and (St) are short calculations with
hereditary substitution, in Sections~\ref{sec:coh-s} and~\ref{sec:coh-d}.

\paragraph{Canonical derivations.} Write $i:B\hookrightarrow\vec B'$ when $\vec B'$
arises from $\vec B$ by inserting $B$ at position $i$. Let
$\mathsf{move}_i:B\to\vec B\Rightarrow X\iso\vec B'\Rightarrow X$ be the derivation
that moves the first argument to position $i$:
$\mathsf{move}_0=\mathrm{id}$ and
$\mathsf{move}_{i+1}=\swap;(\mathrm{id}\to\mathsf{move}_i)$. A hereditary
permutation $h\in\HP(A_1\to\vec A\Rightarrow X,\vec B'\Rightarrow X)$ is a triple
$(i,h_1,h')$ (its \emph{insertion form}): the first argument is inserted at
$i:B\hookrightarrow\vec B'$, $h_1\in\HP(B,A_1)$, and
$h'\in\HP(\vec A\Rightarrow X,\vec B\Rightarrow X)$ permutes the rest. The
\emph{canonical derivation} $\mathsf{der}\,h:A\iso_{\mathrm{ax}}B$ of
$h\in\HP(A,B)$ is defined by recursion on $h$: $\mathsf{der}\,h=\mathrm{id}$ at a
base type, and
\[
\mathsf{der}(i,h_1,h')=\big((\mathsf{der}\,h_1)^{-1}\to\mathsf{der}\,h'\big);\mathsf{move}_i .
\]
This is the derivation of Proposition~\ref{prop:easy}(3); that its meaning is $h$
is (St), proved in Section~\ref{sec:coh-d}.

\subsection{Normal forms}\label{sec:coh-nf}

\begin{lemma}[Insertions commute]\label{lem:inscomm}
  Let $j:C\hookrightarrow\vec D$ insert $C$ into $\vec E$, and
  $j_1:D\hookrightarrow\vec C$ insert $D$ into $\vec D$. Then there are
  $j_1':D\hookrightarrow\vec E'$ (into $\vec E$) and $j':C\hookrightarrow\vec C$
  (into $\vec E'$) with
  \[
  \swap;(\mathrm{id}\to\mathsf{move}_j);\mathsf{move}_{j_1}\;\sim\;
  (\mathrm{id}\to\mathsf{move}_{j_1'});\mathsf{move}_{j'} .
  \]
\end{lemma}
\begin{proof}
  By induction on $j_1$ and $j$. If $j_1=0$, take $j_1'=0$ and $j'=j+1$. If
  $j=0<j_1$, the claim is $\swap;\swap\sim\mathrm{id}$. Otherwise both are
  successors. Then naturality moves $\mathrm{id}\to\mathrm{id}\to\mathsf{move}$
  past a $\swap$, the braid equation applies, and so does the induction
  hypothesis.
\end{proof}

\begin{lemma}[Extraction]\label{lem:extract}
  Let $i:B\hookrightarrow\vec B$ insert $B$ into $\vec B_0$, and
  $k\in\HP(\vec B\Rightarrow X,\vec C\Rightarrow X)$. Then there are
  $j:C\hookrightarrow\vec C$ (into $\vec C_0$), $k_0\in\HP(C,B)$ and
  $k'\in\HP(\vec B_0\Rightarrow X,\vec C_0\Rightarrow X)$ with
  \[
  \mathsf{move}_i;\mathsf{der}\,k\;\sim\;
  \big((\mathsf{der}\,k_0)^{-1}\to\mathsf{der}\,k'\big);\mathsf{move}_j .
  \]
\end{lemma}
\begin{proof}
  By induction on $i$. For $i=0$, take the first insertion of $k$. For $i+1$,
  apply the induction hypothesis inside $\mathrm{id}\to-$, use naturality of
  $\swap$, and then Lemma~\ref{lem:inscomm}.
\end{proof}

\begin{proposition}[Normal forms]\label{prop:nf}
  \leavevmode
  \begin{enumerate}
  \item Let $h\in\HP(A,B)$ and $k\in\HP(B,C)$. Then there is
    $m\in\HP(A,C)$ with $\mathsf{der}\,h;\mathsf{der}\,k\sim\mathsf{der}\,m$.
  \item Every derivation $\varphi:A\iso_{\mathrm{ax}}B$ satisfies
    $\varphi\sim\mathsf{der}\,h$ for some $h\in\HP(A,B)$.
  \end{enumerate}
\end{proposition}
\begin{proof}
  (1) By induction on the height of the types, and then on $h=(i,h_1,h')$. By
  Lemma~\ref{lem:extract},
  \begin{align*}
  \mathsf{der}\,h;\mathsf{der}\,k
  &\sim\big((\mathsf{der}\,h_1)^{-1}\to\mathsf{der}\,h'\big);
     \big((\mathsf{der}\,k_0)^{-1}\to\mathsf{der}\,k'\big);\mathsf{move}_j\\
  &\sim\big((\mathsf{der}\,k_0;\mathsf{der}\,h_1)^{-1}\to
     (\mathsf{der}\,h';\mathsf{der}\,k')\big);\mathsf{move}_j ,
  \end{align*}
  and both composites are canonical by induction: the first at smaller types,
  the second with fewer arguments. Composition alternates between arguments of the
  domain and of the codomain, so the induction is on a bound for the height of all
  three types.

  (2) By induction on $\varphi$, proving the same statement for $\varphi^{-1}$ at
  the same time. Composition uses (1). For $\varphi\to\psi$ the canonical form is
  $(0,h_1,h')$ with $\varphi^{-1}\sim\mathsf{der}\,h_1$ and
  $\psi\sim\mathsf{der}\,h'$; this is why $\varphi^{-1}$ is carried along. For
  $\swap$ it is $(1,\mathrm{id},(0,\mathrm{id},\mathrm{id}))$.
\end{proof}

\paragraph{Remarks.} The equations used in Proposition~\ref{prop:nf} are exactly
those of Section~\ref{sec:setting}: the groupoid laws and functoriality,
$\swap;\swap$ (in Lemma~\ref{lem:inscomm}), naturality (in both lemmas), and the
braid equation (in Lemma~\ref{lem:inscomm}). By (S), each of them holds for
meanings, so the list is sound and, by Theorem~\ref{thm:coh}, complete. The
classical presentation of $S_n$ by Coxeter relations is the special case in
which all $h_1$ are identities: Lemma~\ref{lem:inscomm} is the step in the usual
normal-form proof where a new generator is absorbed, and the hereditary version
needs only naturality in addition.

\subsection{Soundness}\label{sec:coh-s}

In the formalisation the meaning is defined in every context $\Gamma$, because
$\to$-congruence uses it under binders. The meaning at $\Gamma$ is the closed
meaning weakened to $\Gamma$ (by induction on the derivation). So it suffices to
work with closed maps $t\in\Nf(x{:}A,B)$, applied to a term by substitution,
$t\langle u\rangle=t[x=u]$. Composition is $g\langle f\rangle$, and the category
laws are the identity and associativity laws of substitution. Write
$h\Rightarrow k=\lambda f\,a.\,k\langle f\,(h\langle a\rangle)\rangle$. The
remaining equations are computations of the following kind:
\[
(h\Rightarrow k)\langle\lambda a.L\rangle=\lambda a'.\,k\langle
L[a=h\langle a'\rangle]\rangle,\qquad
\swap\langle\lambda a\,b.X\rangle=\lambda b\,a.\,X.
\]
They give functoriality,
$(h_2\Rightarrow k_2)\langle h_1\Rightarrow k_1\rangle=h_1\langle
h_2\rangle\Rightarrow k_2\langle k_1\rangle$, and $\swap\langle\swap\rangle=\mathrm{id}$.
For naturality, both sides compute to
$\lambda f\,b'\,a'.\,c\langle f\,(a\langle a'\rangle)\,(b\langle b'\rangle)\rangle$.
For the braid equation, both sides compute to $\lambda f\,c\,b\,a.\,f\,a\,b\,c$.
The same calculations show that the two parts of the meaning of every derivation
are mutually inverse, by induction on the derivation; this is needed for the
inverse laws.

\subsection{Stability}\label{sec:coh-d}

We prove $\sem{\mathsf{der}\,h}=h$ without computing $\sem{\mathsf{der}\,h}$ by
unfolding hereditary substitution. Instead we compute how meanings act on
argument lists, and recover the term by $\eta$: $t=\lambda\vec y.\,t\,\vec y$. For $i:B\hookrightarrow\vec B'$ and arguments $\vec a$ of types
$\vec B'$, write $a_i$ for the argument at position $i$ and $\vec a\setminus i$ for
the others. Then
\[
\sem{\mathsf{move}_i}\langle F\rangle\,\vec a=F\,a_i\,(\vec a\setminus i),
\]
by induction on $i$, from the actions of $\swap$ and of $\to$:
$\sem{\swap}\langle F\rangle\,b\,a\,\vec s=F\,a\,b\,\vec s$ and
$(h\Rightarrow k)\langle F\rangle\,a\,\vec s=k\langle F\,(h\langle a\rangle)\rangle\,\vec s$.
It follows, by induction on $h=(i,h_1,h')$, that for a variable $x$ and variables
$\vec y$,
\[
\sem{\mathsf{der}\,h}\langle x\,\vec s_0\rangle\,\vec y
=x\,\vec s_0\,\big(\sem{\mathsf{der}\,h_1}\langle y_i\rangle\big)\,
\mathrm{sp}_{h'}(\vec y\setminus i),
\]
where $\mathrm{sp}_{h'}$ is the corresponding spine for $h'$. By induction,
$\sem{\mathsf{der}\,h_1}\langle y_i\rangle=h_1[z=y_i]$, and the spine for $h'$
consists of the arguments of $h'$ with the remaining variables substituted in the
same way. Taking for $x$ the input $a$ with $\vec s_0$ empty, and for $\vec y$
the bound variables, $\eta$ gives
\[
\sem{\mathsf{der}\,h}=\lambda\vec y.\,\sem{\mathsf{der}\,h}\langle a\rangle\,\vec y
=\lambda\vec y.\,a\,\big(h_1[z=y_i]\big)\,\mathrm{sp}_{h'}(\vec y\setminus i),
\]
which is $h$, written in insertion form.

\section{The formalisation}\label{sec:formal}

All results of this paper are formalised in Agda~2.6.5, using only Agda's
builtin modules (no standard library), with no postulates, holes or termination
pragmas. The development has about 5500 lines; the version described here
is tagged \texttt{v2.0} and available at \codeurl.

\paragraph{Syntax.} Types, contexts and de Bruijn variables are as usual; normal
forms are intrinsically typed and in spine form, $\eta$-long by construction
(neutral terms only at base types). Weakening, $\eta$-expansion and hereditary
substitution follow~\cite{ka}, and are accepted by the termination checker without
help. Positions in contexts are telescopes
$(\Gamma\triangleright C)\mathbin{+\!\!+}\Omega$; three rewrite rules for
$\mathbin{+\!\!+}$ (left unit, associativity, and moving a snoc out of a
telescope) make all context equations definitional, and Agda's local confluence
check accepts them. Simultaneous substitution $t[\rho]$ is defined by structural
recursion on $t$, using single-variable hereditary application at a neutral, so it
needs no new termination argument.

\paragraph{Symmetry.} Maps come as pairs with an involution $\mathsf{op}$ that
commutes definitionally with all operations. Every construction and every
one-sided lemma (in particular the parts of Lemma~\ref{lem:descent}) is stated
once and applied in both directions.

\paragraph{Modules.}
\begin{center}
\begin{tabular}{lrp{0.62\textwidth}}
\textbf{module} & \textbf{lines} & \textbf{content}\\\hline
\texttt{Syntax} & 176 & types, normal forms, weakening, $\eta$, hereditary substitution\\
\texttt{Lemmas} & 347 & substitution lemmas; normal forms form a category\\
\texttt{Subst} & 400 & simultaneous substitution, its category laws, pairing\\
\texttt{Iso} & 193 & isomorphisms, derivations and their meaning, list-based $\HP$\\
\texttt{Heads} & 264 & occurrences, free and bound heads; Lemmas~\ref{lem:noncreation}, \ref{lem:heads}\\
\texttt{LamI} & 106 & the $\lambda I$ predicate, weakening and $\eta$\\
\texttt{Tele} & 349 & telescopes, bodies of $\eta$-long terms, spines as substitutions\\
\texttt{Descent} & 305 & matched pairs, Lemma~\ref{lem:descent}, Corollary~\ref{lem:rel}\\
\texttt{Thin} & 168 & thinnings, strengthening, cancellation of renamings\\
\texttt{Count} & 350 & counting occurrences; Lemma~\ref{lem:count}\\
\texttt{Linear} & 102 & Lemma~\ref{lem:linmp}; Lemma~\ref{lem:stray}\\
\texttt{Char} & 271 & Theorem~\ref{thm:char}\\
\texttt{Linearity} & 185 & Corollary~\ref{cor:lin}\\
\texttt{Easy} & 280 & Proposition~\ref{prop:easy}\\
\texttt{Link} & 443 & from the predicate form of $\HP$ to the inductive family\\
\texttt{Coherence} & 394 & level-2 equations, insertion form, Proposition~\ref{prop:nf}\\
\texttt{Sound} & 440 & (S), and all derivations denote isomorphisms\\
\texttt{Actions} & 258 & how meanings act on argument lists; heads, distinct variables\\
\texttt{Derivable} & 140 & (St) for the list form; every isomorphism is the meaning of a derivation\\
\texttt{Stable} & 121 & (St) and faithfulness for insertion forms; Theorem~\ref{thm:coh}\\
\texttt{Examples} & 89 & checks by computation
\end{tabular}
\end{center}

\paragraph{Remarks on the proofs.} Freshness conditions disappear: in a matched
pair, $\alpha_1$ lives over $\Gamma_1$, so by typing it cannot mention the targets.
The descent recurses on accessibility of pairs of types, and the composition of
canonical derivations in Proposition~\ref{prop:nf}(1) on an explicit bound for
the height of all three types. Hereditary permutations are used in three forms: as
a predicate on normal forms (for Theorem~\ref{thm:char}), as an inductive family
with list positions (the statement in \texttt{Iso}), and in insertion form with
an inductive insertion relation (for Theorem~\ref{thm:coh}), which avoids
equations between lists. The first two are linked in \texttt{Link}. The
insertion form is the output of (N). Injectivity in Theorem~\ref{thm:coh} is
proved as in Section~\ref{sec:coh}, from (S), (N) and (St) for insertion forms,
together with their faithfulness (below); surjectivity
combines (St) for the list form with the characterisation. Both proofs of (St)
use the same action formula of Section~\ref{sec:coh-d}, and recover the term by
$\eta$.

\paragraph{Faithfulness of the representation.}\label{par:faithful} In the
paper, $\HP(A,B)$ is a set of normal terms, so in Section~\ref{sec:coh} two
hereditary permutations with the same meaning are equal by definition. In the
formalisation, hereditary permutations are an inductive family of insertion
forms, interpreted as terms by a function $\sem{-}_{\HP}$, and (St) reads
$\sem{\mathsf{der}\,h}=\sem{h}_{\HP}$. So we must also show that this
representation is faithful: insertion forms with the same meaning are equal. This
is a statement about terms only. In
$\sem{h}_{\HP}=\lambda\vec y.\,a\,(\sem{h_1}_{\HP}[z=y_i])\ldots$ the bound
variables $\vec y$ are distinct, and the head of $\sem{h_1}_{\HP}[z=y_i]$ is
$y_i$, so comparing heads fixes the insertion position $i$. Renaming by a single
variable is injective, so induction applies to $h_1$ and to the rest.

\section{Related work}\label{sec:related}

\paragraph{Invertible terms.} Dezani~\cite{dezani} showed that a normalisable
untyped $\lambda$-term is invertible modulo $\beta\eta$ iff it is a finite
hereditary permutation; Bergstra and Klop~\cite{bk} characterised invertible terms
in general. Bucciarelli, De~Faveri, Manzonetto and Salibra~\cite{bdms} recently
showed that the finite hereditary permutations are the invertible terms in every
$\lambda$-theory between $\beta\eta$ and $\mathcal H^+$. Every finite hereditary
permutation is typable, so the typed characterisation follows from the untyped
one~\cite{bdl}. Hereditary permutations are linear terms, so linearity of
isomorphisms is implicit in these results; here it is proved first, from the two
inverse equations, and the characterisation is derived from it.

\paragraph{Isomorphisms of types.} Bruce and Longo~\cite{bl} and Bruce, Di Cosmo
and Longo~\cite{bdl} showed that $\swap$ and congruence axiomatise isomorphism
for $\to$, and extended this to $\times$ and $\one$. Soloviev~\cite{soloviev}
showed, using the model of finite sets, that the corresponding axioms are complete
for all cartesian closed categories. Di~Cosmo's book~\cite{dicosmo} surveys the area, including decision
procedures and applications to library search. With sums and the empty type,
Fiore, Di Cosmo and Balat~\cite{fdb} showed that isomorphism is not finitely
axiomatisable; Ilik~\cite{ilik} studied decidability in this setting. Laurent
characterised isomorphisms through game semantics~\cite{laurent}.

\paragraph{Retractions.} One-sided inverses are much less constrained; see
Regnier and Urzyczyn~\cite{ru} and the references there. Remark~\ref{rem:retract}
shows the phenomena that make one equation insufficient: the head of a section,
relevance and no copying.

\paragraph{Coherence.} The results above are about which types are isomorphic and
which terms are isomorphisms. To our knowledge they do not address when two
derivations denote the same isomorphism; Di~Cosmo's book~\cite{dicosmo} and
survey~\cite{dicosmo-survey} do not treat this question. The classical coherence
theorems for closed categories answer a different one: Kelly and Mac
Lane~\cite{kml} and Mints~\cite{mints77} for (symmetric, monoidal) closed
categories, and Babaev and Solov'ev~\cite{babaev} and Mints~\cite{mints92} for
cartesian closed categories (see Zeilberger~\cite{zeilberger} for a modern
account of Mints' result), show that all
canonical maps between two objects are equal when the objects are
\emph{balanced}, that is, when each atom occurs at most twice. For a balanced type
every automorphism is trivial, so these theorems give no information about
level~2. Theorem~\ref{thm:coh} covers arbitrary types, where the automorphism
groups are non-trivial, and gives a complete set of equations. For symmetric monoidal and symmetric rig
categories the corresponding questions are classical coherence problems (Mac Lane,
Laplaza~\cite{laplaza}); for the reversible language $\Pi$, with sums and
products but no function types, Choudhury, Karwowski and Sabry~\cite{cks} gave a
complete set of level-2 equations, formalised in Agda. Group-theoretically, the
automorphism group of a type tree is an iterated wreath product of symmetric
groups, and Theorem~\ref{thm:coh} gives a presentation of these groups, and of the
groupoid of all isomorphisms, read off the syntax of derivations.

\paragraph{What is new.} (1) A direct typed proof of the characterisation that
does not pass through the untyped calculus, isolates the lemmas needed, and uses
no completeness theorem; the classical alternatives use the untyped
analysis~\cite{dezani}, finite sets~\cite{soloviev}, games~\cite{laurent}, or
Friedman's theorem~\cite{friedman}. (2) The coherence theorem for $\to$.
(3) A complete mechanisation; we have not found an earlier mechanisation of
Dezani's theorem or of its typed version.

\section{From simple types to University Algebra}\label{sec:ua}

The proof was designed to be reused for the syntactic completeness conjecture of
University Algebra. We indicate what we expect to carry over and what changes.

\paragraph{Normal forms and substitution.} For $\to$, $\beta$-normal $\eta$-long
forms with hereditary substitution are canonical. For $\UA$ the type theory has
$\Sigma$, $\Pi$, $\one$ and an extensional boolean $\two$ with a large
eliminator, so equality of terms involves strong $\eta$ for $\two$, as for sums in
the simply typed case~\cite{adhs,scherer}. We expect normal forms in which case
splits on neutral booleans are represented as finite maps, so that the order of
splits does not matter.

\paragraph{The descent.} The analysis of heads should carry over, with one new
feature: a head may sit below case splits, so the permutation $\pi$ becomes a
permutation \emph{per branch}. Relevance and no copying have to be read per
branch too, because erasure and copying in $\UA$ are additive (a branch of a split
may ignore a variable that another branch uses, and different branches each use
it once). The two inverse equations should again interact at the matching of
heads and in no copying, so the counting argument of Lemma~\ref{lem:linmp} has to
be done per branch.

\paragraph{Derivations and level 2.} The role of $\swap$ is taken by the basic
isomorphisms of $\UA$, and the hereditary permutations become hereditary
combinations of permutations, case splits, relabelling of the partitions defined by
boolean-valued functions, and reindexing of domains along automorphisms. The
level-2 equations will include the groupoid laws, functoriality of $\Sigma$ and
$\Pi$, naturality of the basic isomorphisms, and coherence equations of the kind
known for polynomial functors~\cite{gk} and for symmetric rig categories
\cite{laplaza,cks}. The Coxeter presentation of $S_n$ is replaced by presentations
in which only independent entries commute, closer to trace monoids~\cite{diekert}.
The structure of the proof of Theorem~\ref{thm:coh}---soundness, normal forms
without substitution, and stability by computing actions on arguments---is
independent of these details.

\paragraph{An intermediate step.} Between simple types and $\UA$ lies the
logical framework LF: $\Pi$-types over base type families. Hereditary substitution
still terminates by recursion on the simple type obtained by erasing
dependencies~\cite{watkins,hl}, so the presentation of this paper should carry
over, with swaps allowed only between independent arguments. We are pursuing an
intrinsic formalisation of LF along these lines.

\section{Conclusion}

The characterisation of isomorphisms of simple types and its coherence form can be
proved purely syntactically, by an analysis of normal forms that never appeals to
a semantic completeness theorem, and the proof can be formalised directly on
intrinsically typed syntax. The two parts of the coherence proof have very
different natures: the normal-form argument is pure combinatorics on derivations,
while soundness and stability are short calculations with hereditary
substitution. This separation, and the descent on matched pairs, are what we
intend to reuse for University Algebra.

\paragraph{Acknowledgements and use of AI tools.} This paper and its Agda
formalisation were developed in collaboration with Claude (Anthropic), used
through Claude Code: Claude drafted Agda code and text, ran experiments and
searched the literature, under the author's direction. The author has checked the
content and is responsible for it.


\end{document}